\documentclass{aastex}
\usepackage{emulateapj5}

\newcommand\cR{{\cal R}}
\newcommand\km{{\rm \ km}}
\newcommand\simlt{\lower.5ex\hbox{$\; \buildrel < \over \sim \;$}}
\newcommand\simgt{\lower.5ex\hbox{$\; \buildrel > \over \sim \;$}}

\begin{document}

\title{Hyper- and suspended-accretion states of rotating black holes and the
durations of gamma-ray bursts}

\author{Maurice H.P.M. van Putten}
\affil{Department of Mathematics, MIT, Cambridge, MA 02139-4307}
\author{Eve C. Ostriker}
\affil{Department of Astronomy, The University of Maryland, College Park
MD 20742-2421}
\email{mvp@schauder.mit.edu, ostriker@astro.umd.edu}

\begin{abstract}
  We analyze the temporal evolution of accretion onto rotating
  black holes subject to large-scale magnetic torques.
  Wind torques alone drive a disk towards collapse in a finite time
  $\sim t_{ff} E_k/E_B$, where $t_{ff}$ is the
  initial free-fall time and $E_k/E_B$ is the ratio of
  kinetic-to-poloidal-magnetic energy. 
  Additional spin-up torques from a rapidly rotating black hole can arrest
  the disk's inflow. We associate short/long gamma-ray bursts with 
  hyperaccretion/suspended-accretion onto slowly/rapidly spinning black holes.
  This model predicts afterglow emission from short bursts, and may be 
  tested by HETE-II.
\end{abstract}


\section{Introduction}

Active black holes are believed to be surrounded by magnetized disks or tori. 
In particular, such systems are the most 
favorable inner engines of cosmological gamma-ray bursts.
The BATSE catalogue shows a broad, bimodal distribution of GRB
durations ($T_{90}$) 
centered on short bursts of $\sim 0.3\,{\rm s}$ and long bursts of
$\sim 30\, {\rm s}$  \citep{kou93,pac99}.
The cosmological origin of GRBs and their rapid
variability indicate an association with
stellar-mass progenitors \citep{woo93,pac91,pac97,pac98,fry99}.
Proposed scenarios include
coalescence of compact object
binaries \citep{pac91}, and core-collapse
of massive stars \citep{woo93}.
Both scenarios involve a black hole-disk or -torus system.
This disk will be magnetized, due to seeding from the
progenitor's magnetic flux and disk dynamo amplification
\citep{nar92,pac97,bra95,haw96,haw00}.  Various lines of observational
evidence support the collapsar scenario, at least 
for long bursts (e.g. \cite{mvp01}); here, we shall suggest that both
short and long bursts can be explained in a unified model with 
durations extended by coupling to a high-spin black hole.

Accretion provides a definite power
source for GRBs, but the black hole's spin energy, e.g., $2M_\odot c^2$ for
a $7M_\odot$ black hole, can
be at least as important.  Efficient spin energy extraction 
requires an external magnetic 
field supported by 
surrounding matter.
Magnetically mediated energy extraction mechanisms include radiation in
DC Alfv\'en waves (\cite{bla77,tho86}; see also e.g.
\cite{lee99,bes00,kro00}),
or superradiant scattering of fast magnetosonic
waves \citep{uch97,mvp99}.

The evolution of the disk or torus surrounding a black hole depends
crucially on magnetic torques. Negative torques are expected from 
magnetic winds, which can dramatically stimulate accretion rates
(e.g., \cite{bla82,lee00}).
Positive torques $T=-\dot{J}_H$ may derive
from coupling to the black hole's angular momentum $J_H$
by equivalence in poloidal topology to pulsar magnetospheres
(\cite{mvp99,bro01}), and may be key to 
driving hypernovae \citep{bro00}.
The disk may also
redistribute its mass and angular momentum via internal, turbulent,
magnetic and Reynolds stresses \citep{bal98}.  The relative contributions
from {\it external} (wind/black hole) stresses {\it vs.} {\it internal}
stresses is presently unknown, and may depend crucially
on the net poloidal flux introduced to the disk (see e.g. \cite{mat99}).

In this work, we focus on the duration
of magnetized accretion regulated by external torques.
We show that wind torques alone drive the disk 
to a finite-time collapse singularity.  The additional
positive torque from the black hole, when its angular velocity exceeds
a critical value, may arrest accretion onto the black hole for the
duration of its spin-down lifetime.
This introduces two cases: 
a short-lived state of ``hyperaccretion'' for low-spin black holes,
and a long-lived state of ``suspended-accretion'' for high-spin black holes.
The respective time-scales are consistent
with the bimodal distribution in GRB durations for plausible values of 
two presently-uncertain ratios:
the kinetic-to-poloidal-magnetic energy in the disk and the
black hole-to-disk mass.  Numerical simulations soon
within reach may establish the likely ranges of these ratios.

Previously, bimodality in GRB durations has been attributed to, e.g., 
hydrodynamic time-scales in Newtonian vs. relativistic reverse shocks in 
shells decelerating in the ISM \citep{sar95}, or formation 
of millisecond 
pulsars spinning above or below a gravitationally 
unstable limit \citep{yi98}. 
Previous studies of
Poynting dominated winds in GRBs have 
focused on plasma processes and 
possible connections to gamma-ray emission (e.g. \cite{uso92,uso94}; 
\cite{mes97};\cite{bla98};\cite{lyu01}).

\section{Orbital evolution of rotating, magnetized rings subject to wind
  torques}

Magnetic disk-winds transport angular momentum by Reynolds and
Maxwell stresses, which for outgoing (ingoing) flows
have the same (opposite) sign.
A force-free limit is defined by neglecting Reynolds stresses.
Applied to outgoing winds, Maxwell stresses
$-RB_rB_\phi/4\pi$ provide a lower limit on the disk's accretion rate,
where $B$ denotes the magnetic field with components in spherical
coordinates $(r,\theta,\phi)$, and $R=r\sin\theta$.  The force-free limit
represents pure ``magnetic braking.''

A magnetized disk ring, with
mass $\delta M$ and  radius $\cR$ in quasi-Keplerian rotation at rate
$\Omega=(G M_H /\cR^3)^{1/2}$, evolves as it sheds angular momentum in a wind.
Assuming radially asymptotic flow, the magnetic torque
$\delta\tau_B$ on the ring equals \citep{gol69}:
\begin{equation}\label{torque}
\delta \tau_B = {\Omega\over c} \sin^2\theta \left(
{\partial_\mu A_\phi}\right) \delta A_\phi.
\label{EQN_TB2}
\end{equation}
Here, $\mu=\cos\theta$, the poloidal magnetic field is
$B_r=- r^{-2} \partial_\mu A_\phi$, and
$2\pi \delta A_\phi $
is the magnetic flux through the ring.  The toroidal field
is $B_\phi = -\Omega R B_r/c$
(see the wind asymptotics of \cite{gol69,mic69} for aligned rotators;
for disk winds, see \cite{bla76,lee00}).
The ring evolution equation,
$\delta \tau_B=(1/2) \Omega \cR \dot\cR\delta M$, yields
\begin{equation}\label{radvel}
\cR \dot\cR = {2 \over  c}
\sin^2\theta\left({\partial_\mu A_\phi}\right)
\left({\delta A_\phi \over\delta M}\right).
\end{equation}

The evolution of a collapsing ring depends on its
degree of magnetization and the magnetic wind's geometry.  A
split-monopole geometry (SMG) for the asymptotic poloidal field is a
good
approximation when $|B_r|/|B_\phi|>>1$, true
at the onset of accretion for modest rotation
rates.  As the disk collapses and spins up, however, 
magnetic field line winding produces large toroidal magnetic
fields, and a force-free-toroidal-field geometry (TFG) becomes more
appropriate.  SMG/TFG represent two limiting cases, for
which simple solutions to equation (\ref{radvel}) may be obtained.
\footnote{Although, as shown by \cite{bla82} for
(baryonic) MHD disk winds, field line
focusing may produce nonradial asymptotic poloidal fields, 
cylindrically-collimated MHD winds have been found to have
relatively low asymptotic flow speeds \citep{ost97}.}

{\it Solution for split-monopole geometry ---}
In SMG, we have $A_\phi=A(1-\mu/\mu_0)$, with $\theta_0=\arccos(\mu_0)$
the angle of the innermost streamline with respect to the
pole; the outermost streamline lies on the equator.
With $\varpi\equiv \cR/\cR_0$, for $\cR_0$ the
the initial ring radius, the solution of equation
(\ref{radvel}) describes a {\em finite-time singularity:}
\begin{eqnarray}
\varpi(t; A_\phi)=\left(1- {t/ t_f^{smg}} \right)^{1/2},
\end{eqnarray}
where $t_f^{smg}= {(1/4)} {(\delta J^0 / |\delta \tau_{B}^0|)}$, and
$\delta J^0=\delta M \Omega_0 \cR_0^2$ and $\delta \tau_{B}^0$
denote the initial ring angular momentum and torque.

The initial ring kinetic energy is
$\delta E_k^0 =\Omega_0 \delta J^0/2$.
The energy in the magnetic field of a
(double) wedge-shaped portion $\delta A_\phi$
rooted at $\cR_0$ is
$\delta E_B^0=(1/2)\int B_r^2 d\mu r^2dr=
\delta A_\phi|\partial_\mu A_\phi|/(2\cR_0)
=|\delta \tau_B| c (2\cR_0 \Omega_0)^{-1} (\sin\theta)^{-2}$.  For
split-monopole
geometry,
$\delta E_B^0\rightarrow \delta E_{B,s}^0 \equiv A \delta
A_\phi/(2\cR_0)$.
The initial freefall time is
$t_{ff}=\pi/(\Omega_0 2\sqrt{2})$.
For $R_s\equiv 2 G M_H/c^2$ the
Schwarzschild radius of the black hole,
the collapse time in SMG becomes
\begin{equation}\label{tfmon}
{t_f^{smg}}
={ 1 \over 2 c R_s}{\cR_0^2\over \sin^2\theta}{\delta E_k\over \delta
E_{B,s}}
={t_{ff}/\pi\over \sin^2\theta}\left({\cR_0\over R_s}\right)^{1/2}
{\delta E_k\over \delta E_{B,s}}.
\end{equation}
In the above, superscripts on the energies are dropped since
$\delta E_k^0/\delta E_{B,s}^0 =(\delta M/\delta A_\phi) (G M_H/A)$
is an evolutionary invariant.
Equation (\ref{tfmon}) shows that accretion is more efficient
with increasing $\sin\theta$;
outer-disk rings have magnetic field lines closer to the equator
than inner-disk rings, providing greater moment arms for the magnetic torque.

{\it Solution for toroidal field geometry ---}
Under large magnetic hoop stresses, magnetic flux surfaces 
adjust their latitudes such that the {\it toroidal}
magnetic field in the wind approaches a force-free configuration
with $B_\phi\propto R^{-1}$
(see e.g. the protostellar disk wind in \cite{shu95}).
The magnetosphere is then effectively
current-free except along the pole and the equator, with the poloidal pole
current $I_p$ producing
$B_\phi\approx -2 I_p/(c R)$ elsewhere.
Hence,
$\partial_\mu A_\phi = -r^2 B_r= - 2 I_p [\Omega \sin^2\theta]^{-1}$ so
that
$|\delta \tau_B|=2 I_p \delta A_\phi/c$.  Here, $I_p$ relates to the
total normalized flux $A\equiv \int d\mu\,\partial_\mu A_\phi$
by
$I_p\approx -A\Omega(\cR_{in})/(2 |\ln(\theta_0/2)|)$, where $\cR_{in}$
is
the innermost disk radius.

  From equation (\ref{radvel}), the TFG limit gives
$\Omega \cR \dot\cR = -{(4 I_p/c)} {(\delta A_\phi/\delta M)}.$
Using $\Omega(\cR)=\Omega(\cR_0) \varpi^{-3/2}$
we find the {\em finite-time singularity} solution:
\begin{eqnarray}
\varpi(t; A_\phi)=(1-t/t_f^{tfg})^2,
\end{eqnarray}
where
\begin{equation}\label{tfstrat}
{t_f^{tfg}}= {\delta J^0\over |\delta \tau_B^0|}
=t_{ff} {4{\cal I}_0 \over \pi} {\delta E_k\over \delta E_{B,s}}.
\end{equation}
Here,
${\cal I}_0 \equiv {(c/\sqrt{2})} \cR_0^{-1}
\int d\theta (\Omega \sin\theta)^{-1}$ is
$\sim \left({\cR_{in}^3 / R_s \cR_0^2}\right)^{1/2}
|\ln(\theta_0/2)|.$
Going from $\cR_{in}$ to the outside of the disk at $\cR_d$,
the radial factor in ${\cal I}_0$ decreases from
$(\cR_{in}/R_s)^{1/2}$ (which is order-unity) to
$(\cR_{in}^3/R_s\cR_{d}^2)^{1/2}$ (which is of order
$\cR_{in}/\cR_{d}$);
the product $t_{ff} {\cal I}_0$
varies $\propto \cR_0^{1/2}$.
Note that $\delta E_B^0$ in TFG is
larger than $\delta E_{B,s}^0$ by a factor
$|\partial_\mu A_\phi|/A \sim \Omega(\cR_{in})
[\Omega(\cR_0)\sin^2\theta
 |\ln(\theta_0/2)|]^{-1}$.

For the inner disk, $t_f^{tfg}<<t_f^{smg}$ due to the
geometrical factor $4\theta_0^2 |\ln(\theta_0/2)|<<1$; the enhanced
efficiency arises from squeezing of the poloidal field lines
toward the pole by toroidal hoop stresses.  For the outer
disk, $t_f^{tfg}:t_f^{smg}=4 (\cR_{in}/\cR_{d})^{3/2}
|\ln(\theta_0/2)|$, which is of order unity or less unless
the effective beaming angle $\theta_0$ is extremely small.
Thus, although inner-disk accretion is greatly enhanced when the
magnetic wind becomes more stratified toward the poles, the overall
lifetime set by outer disk accretion is not as sensitive to
these geometric changes.

{\it Dimensional estimates of accretion times ---}
Dimensionally, the minimum inner-disk accretion time, from TFG, is
\begin{equation}\label{tinmin}
t_{in}=0.011 {\rm \, s }\
g(\theta_0)\left({\cR_{in}\over 100\km}\right)^2
\left({M_H\over 10 M_\odot}  \right)^{-1} {\delta E_k\over \delta
E_{B,s}},
\end{equation}
where
$g(\theta_0)=\left(1+{\ln(\theta_0/1^\circ)
/\ln(0.009)}\right)$ and
$\theta_0$ is the angle of the innermost wind field line.  
Correspondingly, the maximum outer-disk accretion time is
\begin{equation}\label{tdmax}
t_{d} =0.057 {\rm \, s}\, \left({\cR_d\over 1000\km}\right)^2
\left({M_H\over 10 M_\odot}  \right)^{-1} {\delta E_k\over \delta
E_{B,s}},
\end{equation}
assuming $\cR_d>>\cR_{in}$ with SMG; otherwise, if $\cR_d\sim \cR_{in}$,
we would use the TFG result, which increases the time by a factor
$\sim 4 |\ln(\theta_0/2)|$ consistent with equation (\ref{tinmin}).
These limits will bracket the interval over which disk accretion
takes place.

\section{Disk accretion and black hole spindown times compared}

A ring spends the greatest portion of its lifetime near its
initial radius; during the subsequent rapid infall
phase, magnetic flux is approximately conserved.
For fiducial estimates, we shall normalize to 
an initial kinetic-to-poloidal
magnetic energy ratio $\sim 100$
for all rings, 
as found from \cite{haw00}'s numerical simulations of 
global, hot, MHD accretion disks.
This energy ratio certainly varies in real systems;  estimates based on
simulations with different physics or initial conditions (e.g. 
heating from resistive dissipation, as in \cite{sto00}; 
net poloidal flux, as in \cite{haw96}) may decrease this
ratio toward unity.  The natural diversity in mean magnetizations may in
part be responsible for the broad distributions of the two duration classes
of GRBs.

{\it Time-scale for short GRBs ---}
Applying $\delta E_k/\delta E_{B,s}=10-100$  in equation
(\ref{tinmin}) or (\ref{tdmax}) (for a disk of size $\simlt 1000\km$),
the accretion time will be $\simlt$ a few seconds --
comparable to that of the 2s event GRB 000310C
\citep{jen01}.
In this scenario,
short bursts represent
hyperaccretion onto a slowly-spinning black hole.

{\it Time-scale for long GRBs ---}
What accounts for the long-duration bursts?  If 
the black hole is initially rapidly spinning, then 
the burst may persist for the time $t_H$ required for angular
momentum extraction
by the Blandford-Znajek (1977) and related processes.
Given a horizon flux $2\pi A_{H}$, the black hole torque for open fields is
${\cal L}_{H}\sim {\Omega_H } A_{H}^2/c$ (see \cite{tho86}).
To estimate $A_H$, assume that the black hole field strength is
comparable to
that from the inner part of the disk. In TFG, we find
$A_{H}\approx A (2 |\ln(\theta_0/2)|)^{-1}$ so that
${\cal L}_{H}\sim \Omega_H A^2 (4 c |\ln(\theta_0/2)|^2)^{-1}$.
For comparison,
the total disk-wind angular momentum luminosity is equal to
$\tau_-\equiv \int d\tau_B$; i.e.
${\cal L}_d^{tfg}=\Omega(\cR_{in}) A^2 (c |\ln(\theta_0/2)|)^{-1}$.
The spindown time
$t_{H}=2 G M_H^2 r_H \Omega_H c^{-2}{\cal L}_{H}^{-1}$.
Here, $r_H=R_s\cos^2(\lambda/2)$ is the radius of the
black hole horizon; $\sin\lambda\equiv a/M_H$ is the
ratio of specific angular momentum to the maximal Kerr value.

Taking $\delta E_k/\delta E_{B,s} \sim G M_H M_d/A^2$,
we obtain
$t_{H}= {(8 r_H/ c)} {(\delta E_k/\delta E_{B,s})
(M_H/M_d)}
|\ln(\theta_0/2)|^2.$
Assuming
$a\sim M$ so $r_H\sim R_s/2$, we find
\begin{equation}\label{thspin}
t_{H} = 88 {\rm \, s }\ \left({M_H\over 10 M_\odot}\right)
\left({M_H/M_d \over 100}\right)
\left({\delta E_k/\delta E_{B,s}\over 100}\right)
g^2(\theta_0),
\label{eqnmass}
\end{equation}
consistent with a bimodal offset of
long- from short-duration bursts for $M_d<<M_H$.  Note that 
$t_H/t_{in} \sim M_H/M_d$, as
magnetic torques on the
disk and black hole are comparable, but $J_H$
exceeds $J$ of the inner disk
by a factor of the mass ratio, assuming comparable specific
angular momenta, as for a maximal-Kerr hole.

We remark that a black hole mass $M_H$ of about $10M_\odot$,
formed promptly in the hypernova
scenario,
is expected in view of the Kerr constraint
$J_H\le GM_H^2/c$.
For prompt collapse of a He-core of radius $R_{He}$, stripped of its
hydrogen envelope and tidally locked with a binary companion of mass $m$
\citep{woo93,pac97,ibe96} with an orbital separation $\xi R_{He}$,
\begin{equation}\label{EQN_MH}
M>M_H\ge M \left({2\over 5}\right)^{3}
\left({c^2 R_{He}\over GM \xi^3 }\right)^{3/2}
\left({1+{m\over M}}\right)^{3/2}
\left({\bar{\rho}\over \rho_c}\right)^2.
\end{equation}
Here, the central density $\rho_c$ is that extrapolated from the
He-mantle,
with $\bar{\rho}$ denoting the mean density.
For canonical values
$m=M=20M_\odot$, $R_{He}=R_\odot$ and $\xi=4$,
a black hole may form promptly provided
the inequality on the left hand-side of equation
(\ref{EQN_MH}) is satisfied; i.e.
$1>1.3\times 10^3 \left({\bar{\rho}/\rho_c}\right)^2 $.
This would be true, 
if e.g. the He core is approximately fit by a polytrope with index $n=3$,
for which the Lane-Emden relationship gives
$\rho_c\sim 50 \bar{\rho}$ \citep{kip90}.
Thus, rapidly rotating black holes formed in prompt collapse
tend to be massive.

\section{Suspended accretion and long GRBs}

Long GRBs rely on
the continuing presence of the torus.
Hyperaccretion
times (\S 2) are far shorter, however.
We speculate that resolution of this paradox 
depends on 
the ability of the black hole to spin up the disk
over interconnecting magnetic field-lines equivalent
in poloidal topology to pulsar magnetospheres (Fig. 2 in \cite{mvp99}).
The black hole-torque on the torus is
(adapted from \cite{tho86})
\begin{eqnarray}
\delta\tau_+\approx\frac{1}{c}(\Omega_H-\Omega)\sin^2\theta
             (-\partial_\mu A_\phi)\delta A_\phi,
\label{EQN_PL}
\end{eqnarray}
for a flux-tube of field-lines between them (counting both
above and below the midplane). The torus now has
an inner face interacting with the
black hole, and an outer face interacting with infinity.
The primary role of the inner interaction is to extend the lifetime of 
the torus by preventing its accretion onto the black hole.

The total black hole torque on the torus, $\tau_+$, depends on the
associated horizon flux $2\pi A_H$.
Approach to the horizon tends
to increase $A_H$; if $2\pi A_H$ becomes
large compared to the open torus flux,
an equilibrium state $\tau_+ \sim \tau_-$ may be reached, and
accretion onto the black hole stalls.  This state is expected to be
stable,
since $\tau_+$ decreases with increasing $\cR$: if the torus gains
(loses) angular momentum, it moves outward (inward) and reduces
(increases) $\tau_+$. Such oscillations may give rise to intermittency.
The black hole spindown proceeds on essentially the 
timescale in equation (\ref{thspin});  the difference is that now the torus
- with its two faces -
mediates the interaction.

More quantitatively, integration of equation (\ref{EQN_PL})
gives the net
torque on the torus by the black hole as
$\tau_+\approx {(\Omega_H-\Omega)} A^2 f_H^2/c,$ while the
wind torque is now $\tau_-\approx \Omega A^2 f_w^2 [c
|\ln(\theta_0/2)|]^{-1}$.
Here, $A$ is now $(2\pi)^{-1}$ times the total torus magnetic flux,
and $f_H$ and $f_w$ are the fractions of $A$ passing
through the inner and outer light surfaces, respectively, so that
$f_H+f_w\sim 1/2- 1$.
The torus, ``sandwiched'' between the black hole and infinity, obeys an
evolution equation
\begin{equation}
\cR\dot\cR \approx {2 A^2\over M_d c}
\left[
f_H^2\left({\Omega_H\over \Omega}-1  \right) - {f_w^2 \over
|\ln(\theta_0/2)|}
\right].
\end{equation}
Accretion may therefore be halted -- or
reversed -- if
$\Omega_H \simgt \Omega[1+ (f_w/f_H)^2 |\ln(\theta_0/2)|^{-1}]$.  Early
on,
$f_H<<1$, and this equation will not be satisfied.  Beginning near
the midpoint of accretion when $f_H\simgt f_w$, and provided that
$\Omega_H$ is initiated close to its maximal value, accretion may
be suspended for an interval $\sim t_H$, until the value of $\Omega_H$
drops below the critical value.
Since in the ``suspended" state, the torus gains and loses almost equal
quantities of angular momentum, significant
baryonic matter may be carried off in the wind.

\section{Discussion and comparison with observations}

We have revisited accretion timescales in magnetized black
hole plus disk/torus systems, analyzing
evolution when external torques rather than 
internal ``turbulent'' torques dominate.  We propose two dynamical 
states -- hyperaccretion and suspended accretion -- for systems with
low- or high-spin black holes.  Around a slowly spinning black hole,
wind torques drive the disk to a finite-time singularity -- 
producing short bursts.  Rapidly spinning black holes transfer 
sufficient angular momentum to the disk to arrest matter inflow 
until the black hole spins down -- producing long bursts.
A bimodal duration distribution occurs, provided that $M_d<<M_H$.

Our principal assumption is that external
ordered torques dominate internal disordered ones; this requires
significant poloidal magnetic flux.  Determining how poloidal flux is created 
and when external torques prevail is a major unsolved problem.
Numerical studies to date have shown the development of
magnetic power spectra dominated by the largest scales permitted, in
local disk simulations (e.g. \cite{haw95}), and that
magnetically-driven winds may develop from both cold and hot disks
(e.g. \cite{shi86,sto94,sto00}), with dependence of the angular
momentum loss rate on $E_B$ similar to analytic predictions
\citep{kud98}.  Because the choice of numerical boundary conditions
(particularly for $B_\phi$) may affect the solution at the largest
scale (see e.g. \cite{ust99}), very large dynamic range 
would be needed to assess the external magnetic torque.

Our model differs from expectations for a turbulence-dominated disk in
several ways.  First, a wind-dominated system converts gravitational energy
into Poynting flux rather than thermalizing it.  Such a ``clean'' system
could have less radiative contamination/spectral 
degradation from a low-energy thermal component.  Second, a magnetized wind
is naturally collimated, unlike the fireball in
the alternative scenario.  Finally, only if significant poloidal flux is 
present can field geometry become unfavorable for matter inflow, allowing 
an extended disk lifetime.  In turbulent disks, magnetic fields 
lie primarily parallel to the midplane, so flow onto
the black hole proceeds even as magnetic stresses
transfer spin energy and angular momentum outward \citep{kro99,gam99}.

Because most of the energy is liberated in the innermost regions, 
our hyperaccretion/short GRB and suspended-accretion/long
GRB proposal implies similar intrinsic luminosities, 
but larger fluences in the second case.
We thus
expect short GRBs to feature afterglows similar to those of 
long GRBs, unless
the environments of high- and low- angular momentum 
progenitors differ
appreciably. This may be tested by the HETE-II mission, and is already
supported 
by the afterglow detection to 
the 2s event GRB 000301C \citep{jen01}.

The iron emission lines from
GRB 970508 \citep{piro99} may reflect the composition of progenitor matter.
In the suspended accretion (long GRB) state,
some torus matter is expected to be blown to infinity by the
powerful black hole torque. We associate this baryonic wind with
line-emitting regions for long burst systems; they are
expected to be less powerful in short GRBs.

\acknowledgements 

MVP acknowledges support from
NASA Grant 5-7012 and an MIT C.E. Reed Fund. The authors thank the
ITP/UCSB, where this work was initiated;
NSF supports ITP under PHY94-07194.
We are grateful to  S. Kulkarni, J. Stone, and J. Hawley
for useful discussions, and
thank the referees for constructive comments.

\end{document}